\documentclass[aip,apl,twocolumn,preprintnumbers,amsmath,amssymb,superscriptaddress]{revtex4}
\usepackage{dcolumn}
\usepackage{bm}
\usepackage[dvips]{graphicx}

\graphicspath{{images/}}

\begin{document}

\title{Long range spin supercurrents in ferromagnetic CrO$_2$ using a multilayer contact structure}
\author{M. S. Anwar}
\affiliation{Kamerlingh Onnes Laboratory, Leiden University, P.O.Box 9504, 2300
RA Leiden, The Netherlands}
\author{M. Veldhorst}
\affiliation{Faculty of Science and Technology and MESAþ Institute for
Nanotechnology, University of Twente, 7500 AE Enschede, The Netherlands}
\author{A. Brinkman}
\affiliation{Faculty of Science and Technology and MESAþ Institute for
Nanotechnology, University of Twente, 7500 AE Enschede, The Netherlands}
\author{J. Aarts,$^1$}
%
\date{Date of submission \today}

\begin{abstract}
We report measurements of long ranged supercurrents through ferromagnetic and
fully spin-polarized CrO$_2$ deposited on TiO$_2$ substrates. In earlier work,
we found supercurrents in films grown on sapphire but not on TiO$_2$. Here we
employed a special contact arrangement, consisting of a Ni/Cu sandwich between
the film and the superconducting amorphous Mo$_{70}$Ge$_{30}$ electrodes. The
distance between the contacts was almost a micrometer, and we find the critical
current density to be significantly higher than found in the films deposited on
sapphire. We argue this is due to spin mixing in the Ni/Cu/CrO$_2$ layer
structure, which is helpful in the generation of the odd-frequency spin triplet
correlations needed to carry the supercurrent.

\end{abstract}


\maketitle

\begin{figure}[t]
\begin{center}
\includegraphics[width=8cm]{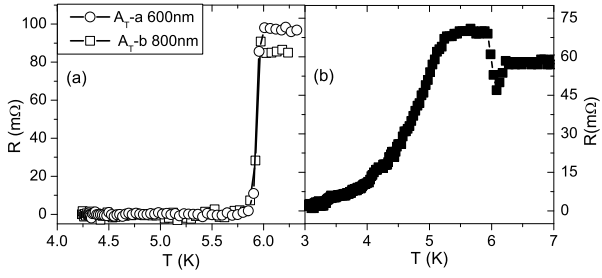}
\caption{Resistance $R$ versus Temperature $T$, (a) for junctions A$_T$-a (gap
600 nm; electrode width 30 $\mu$m) and A$_T$-b (gap 800 nm); (b) for junction
B$_T$ (gap 700 nm, electrode width 5~$\mu$m).} \label{rtmultilayer}
\end{center}
\end{figure}

\begin{figure}[t]
\begin{center}
\includegraphics[width=9cm]{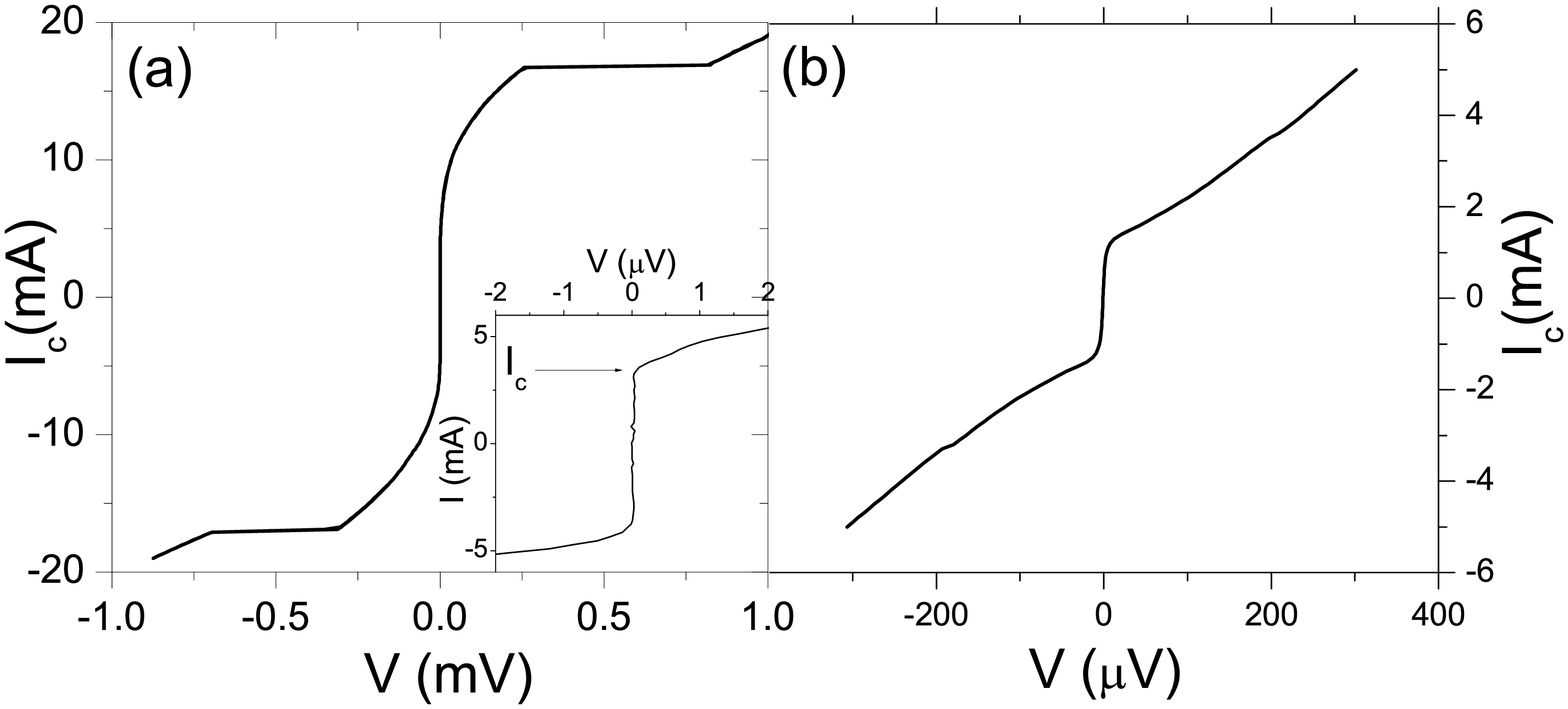}
\caption{Current $I$ versus Voltage $V$ measured (a) for junction A$_T$-b at
4.2~K, and (b) for junction B$_T$ at 3~K.} \label{IVT}
\end{center}
\end{figure}

\begin{figure}[b]
\begin{center}
\includegraphics[width=6cm]{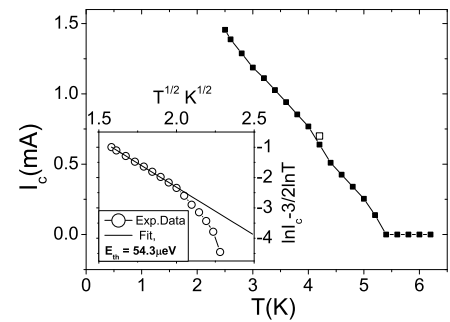}
\caption{$I_c(T)$ for junction B$_T$. The open symbol is $I_c$ at 4.2~K for
junction C$_T$, as follows from Fig.\ref{IVT} Inset: plot of (ln(I$_c$)-(3/2
ln(T)) versus $\sqrt{T}$ to determine the Thouless energy E$_{th}$ =
54~$\mu$eV.} \label{icttio2}
\end{center}
\end{figure}

\begin{figure}[t]
\begin{center}
\includegraphics[width=8cm]{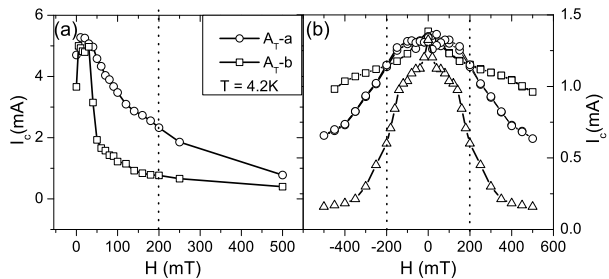}
\caption{Critical current $I_c$ versus applied field $(H_a)$ (a) at 4.2~K for
junctions A$_T$-a ($\bigcirc$) and A$_T$-b ($\Box$) with $H_a$ in-plane and
$\perp$ current $I$; (b) for junction B$_T$ at 3~K in three different
configurations, in-plane $H_a \parallel I$ ($\Box$), $H_a \perp I$
($\bigcirc$), and out-of-plane $H_a \perp I$ ($\triangle$). The vertical dotted
lines indicate the field at 200~mT for reference purposes.} \label{ichtio2}
\end{center}
\end{figure}

Conventional spin-singlet Cooper pairs from a superconductor (S) dephase over a
coherence length $\xi_F=\sqrt{\hbar D_F/h_{ex}}$ (dirty limit) in a ferromagnet
(F) under the influence of its exchange field $h_{ex}$ (and $D_F$ the diffusion
constant in the F-metal). Even for weak ferromagnets, $\xi_F$ is only a few nm.
Such dephasing would not occur with equal-spin triplet Cooper pairs, leading to
a long range proximity (LRP) effect in the ferromagnet. It was predicted that
triplet correlations can be induced at an S/F interface when $h_{ex}$ is
inhomogeneous \cite{Bergeret01,Kadigrobov01,Eschrig08}, for instance from
domain walls or unaligned magnetic moments. This should also allow a Josephson
current in an S/F/S geometry. To observe this, both interfaces are required to
show similar inhomogeneities \cite{Houzet07} as for instance in an
S/F$_1$/F/F$_2$/S trilayer in which the magnetizations of the F$_1$,
F$_2$ layers are non-collinear with the central F layer. \\
Early work on CrO$_2$ \cite{Keizer06} and Holmium \cite{Sosnin06} gave the
first indications for such LRP effects in ferromagnets. In the first case, a
supercurrent was measured in devices where superconducting electrodes of NbTiN
with separations up to 1 $\mu$m were placed on unstructured 100~nm thick films
of CrO$_2$ (a half metallic ferromagnet or HMF) which were grown on TiO$_2$
substrates. In the second case, the LRP effect was observed in ferromagnetic Ho
wires of lengths up to 150~nm using an Andreev interferometer geometry. More
recently, LRP effect were reported using Josephson junctions where a Co central
layer was used in combination with PdNi, CuNi or Ni layers
\cite{Khaire10,Khasawneh11}; and where a Co layer was used together with Ho
layers to provide magnetic inhomogeneity \cite{Robinson10}. Signatures of LRP
effect were also observed with the Heusler Compound Cu$_{2}$MnAl
\cite{Sprungmann10} and in Co nanowires \cite{Wang10}. At the same time, the
observation of supercurrents over a length of 700~nm through
CrO$_2$ deposited on sapphire substrates was reported \cite{Anwar10,Anwar11}. \\
The experiments with Co junctions were up to Co thicknesses of 50~nm. Since Co
is not fully spin polarized the triplet decay is mainly set by the spin
diffusion length, and can be expected to be of the order of 100~nm. That makes
the CrO$_{2}$ case with its significantly larger decay length of special
interest, but in the previous experiments the reproducibility was an issue. In
particular, it was not clear where the inhomogeneous magnetization resides
which is needed for the triplet generation. Also, in our previous work we did
not succeed in finding supercurrents in films deposited on TiO$_2$. Here we
report on observing long ranged supercurrents in CrO$_2$ grown on TiO$_2$,
using 2~nm Ni as an extra layer in the contact geometry to induce an artificial
magnetic inhomogeneity, and 5~nm Cu to magnetically decouple the Ni and the
CrO$_2$. We find much stronger supercurrents than in the case of sapphire,
indicating that with the Ni/Cu sandwich we have a good generator for triplet
Cooper pairs.\\
The devices were fabricated in a lateral geometry using 60~nm thick
a-Mo$_{70}$Ge$_{30}$ superconducting contacts (transition temperature $T_{c}$ =
6~K) deposited on unstructured 100~nm thick CrO$_{2}$ films grown on TiO$_2$
substrates. We made the devices through a lift-off mask using a bilayer resist.
Ar-ion etching was applied immediately prior to deposition, in order to remove
the Cr$_{2}$O$_{3}$ on the film surface, and the Cu/Ni/Mo$_{70}$Ge$_{30}$
sandwiches were sputtered {\it in situ}. Two junctions were made on each
sample, perpendicular to each other, and both junctions were measured
independently. More details can be found in
Refs.\cite{Anwar10,Anwar11} \\
A supercurrent was measured successfully in three devices out of five, named
A$_T$, B$_T$ and C$_T$. On A$_T$ (30~$\mu$m wide leads) both junctions showed a
supercurrent. We call them A$_T$-a (600 nm gap) and A$_T$-b (800 nm gap).
Samples B$_T$ and C$_T$ were prepared with 5~$\mu$m wide leads, in order to
lower the absolute value of the currents, and a gap of 700~nm. Here only one
junction was showing a measurable critical current on each sample. Sample C$_T$
was measured in two different cryostats, one with extra filtering to minimize
to amplifier contribution to the data in the zero-voltage branch. A drawback
still is the limited lifetime of the samples. The supercurrent disappears after
a few cool-downs, possibly due to the effect of thermal cycling on the films.
\\
For sample A$_T$, the resistance $R$ as function of temperature $T$ is given in
Fig.~\ref{rtmultilayer}a and shows a sharp down-jump at $T_c$. For junction
B$_T$ (Fig.~\ref{rtmultilayer}b), $R(T)$ shows a small dip at 6~K, followed by
an up-jump, a flat part, and then a slow decrease. For junction C$_T$ the
behavior is similar but with a larger up-jump to 0.7~$\Omega$, similar to our
sapphire-based devices \cite{Anwar11}. \\
%
Figure \ref{IVT}a shows an $I$-$V$ characteristic for sample A$_T$-b, measured
at 4.2~K. There is a zero-resistance branch up to a well-defined current of
about 3~mA at which a finite voltage develops. On larger scales a bend in the
curve is seen, followed by another transition at 15~mA to Ohmic behavior with
R$_{N}$=100~m$\Omega$. Figure \ref{IVT}b shows $I$-$V$ data measured on sample
B$_T$ at 3~K. The value for $I_c$ is 1.2~mA, and the resistive branch has a
value of 80~m$\Omega$, in very reasonable agreement with the normal state
resistance. The residual resistance below $I_c$ is a few m$\Omega$. Sample
C$_T$ was first measured at 4.2~K in a cryostat with well-filtered leads. Here,
the $I$-$V$ characteristic showed sharp switching and some hysteretic behavior,
with $I_c$ of the order of 0.5~mA. The residual resistance below $I_c$ is
3~m$\Omega$.
$I_c(T)$ was defined by a 1~$\mu$V criterion and measured for junction B$_T$
and C$_T$ in the temperature range of 2.5 K to 6~K. As shown in
Fig.~\ref{icttio2} for sample B$_T$ the behavior is almost linear. For sample
A$_T$ we first measured the field dependence of $I_c$ at 4.2~K, but we did not
measure $I_c(T)$ because the supercurrent disappeared after the third
cool-down. The measurement on sample C$_T$ is also shown in Fig.~\ref{icttio2}.
In a subsequent measurement, $I_c$ had gone down to 70~$\mu A$, illustrating
the fragility of the sample, but $I_c(T)$ also showed a linear increase.
Figure~\ref{ichtio2}a illustrates the effect of a magnetic field $H_a$ on $I_c$
at 4.2~K for both junctions A$_T$-a and A$_T$-b, with $H_a$ in the plane of the
junction and $\perp I$. It shows that $I_c$ in the case of A$_T$-a is quite
sensitive to $H_a$, with an initial fast decrease below 60~mT, but less so in
the case of A$_T$-b. Figure~\ref{ichtio2}b presents $I_c(H_a)$ at 3~K for
junction B$_T$ in three different configurations, $H_a$ in-plane and
$\parallel, \perp I$, and $H_a$ out-of-plane. Here the field-in-plane data show
a relatively slow decrease, while the field-out-of-plane data show a small
sharp peak, followed by a shoulder around 100~mT. Neither for A$_T$-a,b nor for
B$_T$ there is evidence for a Fraunhofer pattern.

The claim from the measurements is that large supercurrents are now flowing
through the CrO$_2$ bridge. In discussing these results we address the
following issues. We compare the residual resistance in the supercurrent
measurements with the normal state resistance of the bridge; we discuss the
possibility of depairing currents in the superconducting leads; a Thouless
analysis is performed; and we discuss the effects of applying a magnetic
field. \\
The $I_c$'s measured here can be compared with our previous measurements
\cite{Anwar10,Anwar11} on sapphire-based junctions. The current density at
4.2~K, (d$_{CrO_2}$ $\approx$ 100~nm, junction width 30~$\mu$m and 5~$\mu$m,
current $\approx$ 3~mA and 0.5~mA respectively) is of the order of 1 $\times$
10$^{9}$ A/m$^{2}$ for A$_T$, B$_T$, as well as C$_T$. In all cases, it is 100
times larger than that of sapphire-based junctions, and of similar magnitude as
in the earlier observations of Keizer {\it et al.} \cite{Keizer06}. This
suggests that a uniform spin active interface is present at the interface, due
to the additional 2~nm Ni layer.
\\
An important question is whether the $I-V$ characteristics such as shown in
Fig. \ref{IVT} are truly from the CrO$_2$ bridge, and not just the
superconducting contacts. For this we take another look at the normal
resistance of the bridge. Taking $\rho_{CrO_2}$=10 $\mu\Omega$cm, a film
thickness of 100~nm, a bridge width of 5~$\mu$m, and a junction length of
700~nm, $R_N$ comes out to be 140~m$\Omega$ (25~m$\Omega$ for the 30~$\mu$m
wide contacts). This is significantly higher than what is measured in the
zero-voltage branch of the $I$-$V$ characteristics, where it is not more than a
few m$\Omega$. Note that the measured resistance above $T_c$ is higher than the
above estimate. This is because, when the superconducting leads become normal,
the geometry of the sample is a very different one, with both high resistance
MoGe and low-resistance CrO$_2$ contributing.
\\
Another issue is how close $I_c$ comes to the depairing current $I_{dp}$ of the
superconducting leads. For the sapphire-based junctions with their low $I_c$
values this was not relevant. The value for $J_{dp}$ of a-MoGe is about 4
$\times$ 10$^{10}$~A/m$^2$ at 4.2~K \cite{Rusanov04}. Taking into account that
the thickness of the lead (40~nm) is smaller than that of the bridge, the
current density in the lead at the measured $I_c$ for all junctions is about
$2.5 \times 10^9$ A/m$^2$, still an order of magnitude smaller than $J_{dp}$.
This probably explains, however, the second transition seen in Fig.\ref{IVT}a,
which takes place at a 5 times higher current density.
\\
Although $I_c(T)$ is quite linear, the Thouless energy of the junction can be
estimated from a plot of (ln($I_c$)-(3/2 ln(T)) versus $\sqrt{T}$ (see inset of
Fig.\ref{icttio2}). For junction B$_T$ we find E$_{th}$ = 54~$\mu$eV, not much
different from that of sapphire based junctions \cite{Anwar10,Anwar11}. From
E$_{Th}$ we can estimate $I_c$ at 4.2~K using theoretical results for a long
junction \cite{dubos01}. For (k$_B T$/E$_{Th}$) $\approx$ 7.6, we find from
Ref.\cite{dubos01} that $I_c R_N$ $\approx$ $E_{Th}$ $\approx$ 54 $\mu$V, which
with $R_N$ = 60~m$\Omega$ leads to $I_c$ = 0.9~mA , quite close to the measured
value.
\\
The magnetic field effects are complicated. For $H_a \parallel I$ the junctions
A$_T$-a,b are more sensitive to the field than B$_T$. For junctions A$_T$-a,b
the first sharp decrease at 60~mT might correspond to the first flux quantum,
which is a reasonable value according to the dimensions of the junctions, but
no such behavior is seen for B$_T$. The suppression of $I_c$ is stronger than
in the earlier work. Taking a 200~mT field as a reference point, the
suppression is over 70\% for A$_T$, and still almost 30\% for B$_T$, compared
to 10\% in the sapphire-based junctions. This points to a diminishing
effectiveness of the Ni/Cu layer, although it might be argued that the effect
should be even stronger: in 200~mT both the CrO$_2$ and Ni magnetization should
be saturated and aligned, removing a possible source of magnetic inhomogeneity.
Instead, the supercurrents were not even quenched in 500~mT. It suggests that
there is a residual magnetic inhomogeneity residing in the Ni/Cu/CrO$_2$
sandwich, which is not removed by the magnetic field. This needs further study.
\\
In conclusion, a Ni/Cu sandwich on top of ferromagnetic CrO$_2$ deposited on
TiO$_2$ substrates leads to strong supercurrents over a distance of almost
1~$\mu$m. The Ni/Cu sandwich appears to furnish spin mixing and triplet
generation similar to what was found in Co-based junctions.
\\
\indent This work is part of the research program of the Stichting F.O.M.,
which is financially supported by NWO. M.S.A. acknowledges the financial
support of the Higher Education Commission (HEC) Pakistan.

\end{document}